# E2 and M1 strengths in heavy deformed nuclei [*]

J. P. Draayer [†], G. Popa [‡]

Department of Physics and Astronomy, Louisiana State University,
Baton Rouge, LA 70803-4001, USA

AND

J. G. Hirsch [§]

Instituto de Ciencias Nucleares, Universidad Nacional Autónoma de México,
Apartado Postal 70-543 México 04510 DF, México

Energy levels of the four lowest bands in $^{160,162,164}$Dy and $^{168}$Er, B(E2) transition strengths between the levels, and the B(M1) strength distribution of the ground state, all calculated within the framework of pseudo-$SU_3$ model, are presented [9]. Realistic single-particle energies and quadrupole-quadrupole and pairing interaction strengths fixed from systematics were used in the calculations [6]. The strengths of four rotor-like terms, all small relative to the other terms in the interaction, were adjusted to give an overall best fit to the energy spectra. The procedure yielded consistent parameter sets for the four nuclei.

## 1. Introduction

Experimental nuclear physicists, with improved equipment and techniques, continue to challenge nuclear theorists with interesting new phenomena. The measurement of new levels, some lying below 3 MeV, raises questions about the nature of collective excitations in atomic nuclei. Even though heavy deformed nuclei with $A \geq 150$ are good candidates for probing such degrees of freedom, microscopic calculations are very important to gaining a deeper understanding of the nature, for example, of low-lying

---

excited $0^+$ bands and $K^\pi = 4^+$ states. Of special interest to us is the fragmentation of the ground state $M1$ strength distribution.

Over the last several years, various properties of low-energy states in $^{160}$Dy [1], $^{162}$Dy [2], $^{164}$Dy [3] and $^{168}$Er [3, 4, 5] as well as other rare earth nuclei have been measured. In a recent article, we presented a description of the even-even $^{156,158,160}$Gd isotopes [6] within the framework of the pseudo-$SU_3$ model. One of our main goals of the present study was to continue to test the Hamiltonian and obtain a consistent set of parameters for a larger set of nuclei. In order to achieve this goal we fixed the strengths of the dominant terms in the interaction: single-particle energies as well as the two-body quadrupole-quadrupole and pairing interactions. Four smaller terms were varied to "fine tune" the results to experiment.

In this paper we apply the pseudo-$SU_3$ model to the normal parity bands in the $^{160,162,164}$Dy and $^{168}$Er. These nuclei, as for the Gd isotopes studied earlier, all exhibit well-developed rotational ground state bands as well as states that are associated with low-lying $K^\pi = 0^+$ and $K^\pi = 2^+$ bands. Here we provide a theoretical description for states of the low-lying bands in each of these isotopes including B(E2) transition strengths between the states and the ground state B(M1) strength distribution. As appropriate, we also give results for the $K^\pi = 4^+$ and $K^\pi = 1^+$ bands in some of these nuclei. The theory yields a reasonable reproduction of the observed low-lying $1^+$ states, the ground state $M1$ sumrule and its energy-weighted centroid, as well as the observed fragmentation of the $M1$ strength but not to the same level of accuracy as for other properties.

The Hamiltonian we used in the study is given in Section 2. Excitation energies, intra-band B(E2) transition strengths, and ground state B(M1) strength distributions are reported in Section 3. Some conclusions that follow from the analyses are offered in Section 4.

## 2. Model space and parameters

The nuclei considered in the present calculations have closed shells at $N_\pi = 50$ for protons and $N_\nu = 82$ for neutrons. To build basis states, we considered an open oscillator shell for each ($\eta_\pi = 4$ for protons and $\eta_\nu = 5$ for neutrons) along with their intruder state complements ($h_{11/2}$ for protons and $i_{13/2}$ for neutrons) even though particles in these unique-parity intruder levels were only considered to renormalize the normal-parity configurations through the use of an effective charge. These oscillator shells have a complementary pseudo-harmonic oscillator shell structure given by $\tilde{\eta}_\sigma$ ($\sigma = \pi, \nu$) $= \eta_\sigma - 1$. Approximately 20 pseudo-$SU_3$ irreducible representation (irreps) with largest values of the second order Casimir operator $C_2$ ($Q \cdot Q = 4C_2 - 3L^2$) were used in building the basis states.

We used the following realistic pseudo-$SU_3$ Hamiltonian:

$$H = H^{\pi}_{sp} + H^{\nu}_{sp} - \frac{1}{2}\chi\, Q \cdot Q - G_{\pi}\, H^{\pi}_P - G_{\nu}\, H^{\nu}_P$$
$$+ a\, J^2 + b\, K_J^2 + a_3\, C_3 + a_{sym}\, C_2\, . \tag{1}$$

Strengths of the quadrupole-quadrupole ($Q \cdot Q$) and pairing interactions ($H^{\sigma}_P$) were fixed, respectively, at values typical of those used by other authors, namely, $\chi = 35\, A^{5/3}$ MeV, $G_{\pi} = 21/A$ MeV and $G_{\nu} = 19/A$ MeV. The spherical single-particle terms in this expression have the form

$$H^{\sigma}_{sp} = \sum_{i_{\sigma}}(C_{\sigma}\, \mathbf{l}_{i_{\sigma}} \cdot \mathbf{s}_{i_{\sigma}} + D_{\sigma}\, \mathbf{l}^2_{i_{\sigma}})\, . \tag{2}$$

Since only pseudo-spin zero states were considered, matrix elements of the spin-orbit part of this interaction vanish identically. Calculations were carried out under the assumption that the single-particle orbit-orbit ($l^2$) interaction strengths were fixed by systematics [7],

$$D_{\sigma}(\sigma = \pi, \nu) = \hbar\omega\kappa_{\sigma}\mu_{\sigma}\, , \tag{3}$$

where $\hbar\omega = 41/A^{1/3}$ with $\kappa_{\sigma}$ and $\mu_{\sigma}$ assigned their usual oscillator values [7]:

$$\kappa_{\pi} = 0.0637,\quad \mu_{\pi} = 0.60$$
$$\kappa_{\nu} = 0.0637,\quad \mu_{\nu} = 0.42\, . \tag{4}$$

| parameter | $^{168}$Er | $^{164}$Dy | $^{162}$Dy | $^{160}$Dy |
|---|---|---|---|---|
| $\hbar\omega$ | 7.40 | 7.49 | 7.52 | 7.55 |
| $\chi \times 10^{-3}$ | 6.84 | 7.12 | 7.27 | 7.42 |
| $D_{\pi}$ | -0.283 | -0.286 | -0.287 | -0.289 |
| $D_{\nu}$ | -0.198 | -0.200 | -0.201 | -0.202 |
| $G_{\pi}$ | 0.125 | 0.128 | 0.130 | 0.131 |
| $G_{\nu}$ | 0.101 | 0.104 | 0.105 | 0.106 |
| a $\times 10^{-3}$ | -2.1 | -2.0 | 0.0 | 1.0 |
| b | 0.022 | 0.00 | 0.08 | 0.10 |
| $a_{sym} \times 10^{-3}$ | 0.80 | 1.20 | 1.40 | 1.45 |
| $a_3 \times 10^{-4}$ | 0.75 | 0.65 | 1.32 | 1.36 |

Table 1. Parameters of the pseudo-$SU_3$ Hamiltonian.

Relative excitation energies for states with angular momentum $0^+$ are determined mainly by the quadrupole-quadrupole interaction. The single-particle terms and pairing interactions mix these states. With the strength of these interactions fixed as in Table 1, the $0_2^+$ states lie very close to their experimental counterparts while the $0_3^+$ states usually slightly above the experimental ones. Of the four "free" parameters in the Hamiltonian, $a$ was adjusted to reproduce the moment of inertia of the ground state band, $a_3$ was varied to yield a best fit to the energy of the second $0^+$ state (the energy of the third $0^+$ was not included in the fitting and as the results given below show these all fall slightly higher than their experimental counterparts), $a_{sym}$ was adjusted to give a best fit to the first $1^+$ state, and $b$ was fit to the value of the band-head energy of the $K^\pi = 2^+$ band.

In the rotational model the projection $K$ of angular momentum on the body-fixed symmetry axis is a good quantum number. For each intrinsic state with a given value of $K$ there is a set of levels with $L = K, \ K+1, \ K+2, \ldots$, except for $K = 0$ when $L$ is either even or odd depending on the intrinsic $(D_2)$ symmetry of the configuration. Elliott [8] used group-theoretical methods to investigate classification schemes for particles in a three-dimensional harmonic oscillator potential for which the underlying symmetry is $SU_3$. He noted that the angular momenta in an irrep of $SU_3$ can be grouped in a similar way to that of the rotor, the differences being that there are a fixed number of $K$ values and that each band supports a finite number of $L$ values rather than being of infinite length. The angular momentum content of an $SU_3$ irrep $(\lambda, \mu)$ can be sorted into $K$ bands according to the following rule [9]:

$$K = \min(\lambda, \mu), \ \min(\lambda, \mu) - 2, \ \ldots, \ 1 \text{ or } 0, \tag{5}$$

where

$$L = (\lambda + \mu), \ (\lambda + \mu) - 2, \ \ldots, \ 1 \text{ or } 0 \tag{6}$$

for $K = 0$ and

$$L = K, \ K+1, \ K+2, \ \ldots, (\lambda + \mu) - K + 1 \tag{7}$$

for $K \neq 0$. Hence, the allowed angular momentum values in the leading $SU_3$ irrep for $^{160}$Dy with (28,8) are $L = 0, \ 2, ..., 36$ for the $K = 0$ band, $L = 2, \ 3, ..., 35$ for the $K = 2$ band, etc.

## 3. Results

Using the Hamiltonian and procedure described in the previous section, good agreement was obtained between the experimental and calculated energies of the first three low-energy bands in each of the nuclei considered. States in the ground-state, first excited $K = 2$ and first excited $K = 0$ bands were all found to lie very close to their experimental counterparts. A second excited $K^\pi = 0^+$ band was identified at approximately 0.5 $MeV$ above its experimental counterpart. As noted above, this level was not included in the fitting procedure. Two other bands, $K^\pi = 1^+$ and $K^\pi = 4^+$, were also identified for each of the nuclei.

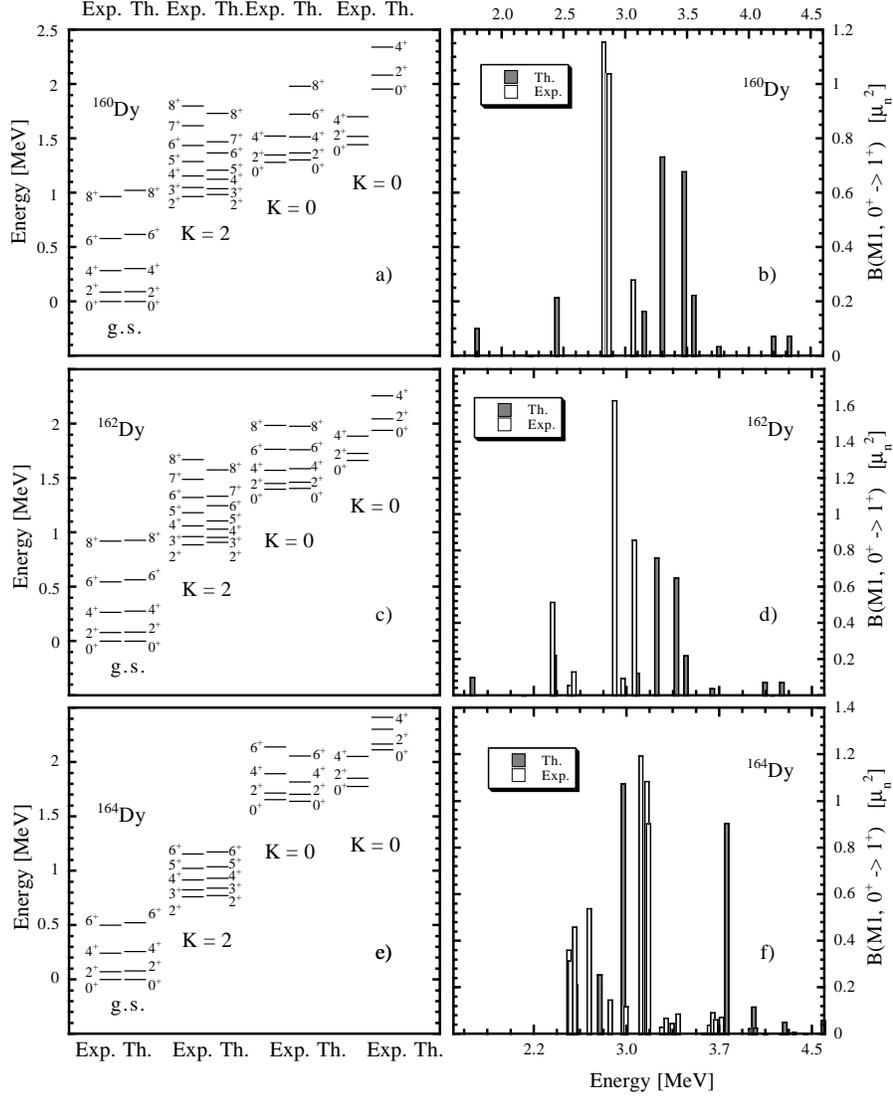

Fig. 1. Energy spectra of $^{160}$Dy, $^{162}$Dy, and $^{164}$Dy obtained using Hamiltonian 1. 'Exp.' represents the experimental results and 'Th.' the calculated ones. Figures b), d), and f) give the theoretical and experimental $M1$ transition strengths from the $J = 0$ ground state to the various $J = 1$ states.

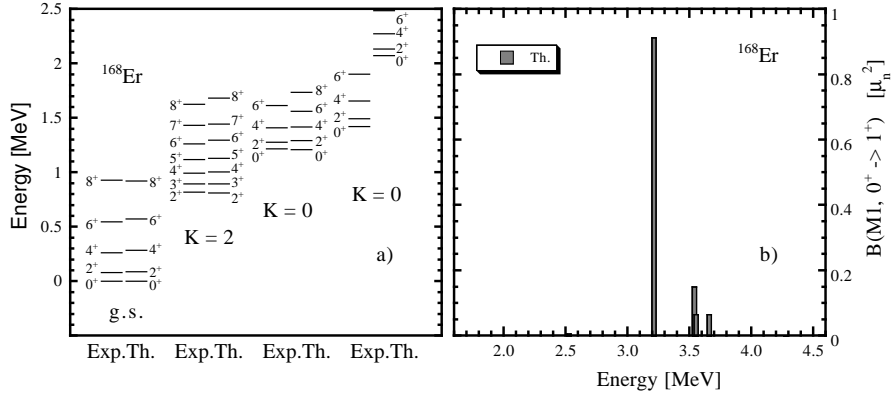

Fig. 2. Energy spectra of $^{168}$Er obtained using Hamiltonian (1). 'Exp.' represents the experimental results and 'Th.' the calculated ones. Figures b) gives the theoretical predictions for $M1$ transition strengths from the $J = 0$ ground state to the various $J = 1$ states.

As can be seen from Table 1, the parameters that the fitting procedure yielded for the Dy isotopes vary smoothly from one nucleus to another. In fact all four of the 'free' parameters decrease as the mass number $A$ increases. Results were also generated for $^{168}$Er which has one pair of protons more than the Dy isotopes. Strengths of the interactions in this case are given in second column of Table 1. Since $A$ is larger for $^{168}$Er than for the Dy isotopes, one might expected even smaller values for the parameters, and while this is what is observed for $a$ and $a_{sym}$, the extra proton pair requires values for $b$ and $a_3$ that are smaller than the ones for $^{160,162}$Dy but larger than those for $^{164}$Dy.

Figure 1(a) shows the calculated and experimental [10] $K = 0$, $K = 2$ as well as the first and second excited $K = 0$ bands for $^{160}$Dy. The agreement between theory and experiment is excellent for the first three bands with relative differences between calculated and experimental energies being in all cases less than 7%. The model predicts a continuation of the first excited $K = 0$ band with two additional states of angular momentum 6 and 8 in $^{160}$Dy. The second excited $K = 0$ band is identified to be about 0.5 $MeV$ higher than the experimental one. As for $^{160}$Dy, the agreement between theory and experiment is excellent for the first three bands in $^{162}$Dy and $^{164}$Dy. Again, the second excited $K = 0$ band is identified at slightly higher than the experimental one. The experimental and calculated energy spectra for these nuclei are given in Figs. 1(c) and 1(e).

The energy spectra for $^{168}$Er is given in Fig. 2 (a). Even though this nucleus has one additional pair of protons, the calculated energy spectra exhibits the same behavior as is Dy isotopes. The ground-state, $K = 2$, and first excited $K = 0$ bands are well reproduced. An angular momentum 8 state is identified as a continuation of the first excited $K = 0$ band. The second excited $K = 0$ band is identified and lies higher than the experimental one. As Fig. 1 shows, the difference in energy between the calculated and experimental band heads of the second excited $K = 0$ bands increases as the mass number $A$ increases. For $^{160}$Dy and $^{162}$Dy the difference is about 0.3 MeV while for $^{164}$Dy it is about 0.5 MeV, and for $^{168}$Er is about 0.65 MeV. The states in the second excited $K = 0$ band are well defined. The results suggest that it may be interesting to see how the parameters change with changing proton rather than neutron number.

Collective models have emerged as a result of attempts to reproduce experimental observations with simple calculations. In the evolution of the $SU_3$ model, and later the pseudo-$SU_3$ model, one of the main motivations was to achieve a good description of deformed nuclei using a small configuration space. In the present study the configuration spaces are small compared with typical ones that are used in shell-model calculations based on m-scheme configurations. To provide a better understanding of the theory it is useful to take a closer look at the eigenvectors of the states in the four bands under consideration. This is done in Table 2 where the $SU_3$ content of the calculated eigenvectors for states of the four lowest bands in $^{162}$Dy are given. The percentage distributions of each eigenvector across the $(\lambda, \mu)$ values are given in the second column. Only basis states that contribute more than 2% are identified. In the ground state band there is clearly a dominant $SU_3$ irrep which is the leading irrep in the coupling of the leading proton and neutron $SU_3$ irreps. An additional five irreps contribute in total with about 40 % to the eigenvector strength. The same leading irrep is the most important in the $K = 2$ band, and as we will see later in the $K = 4$ band. Again there are only about six irreps that contribute more than about 2% to the eigenvectors. The first excited $K = 0$ band exhibits a different dominant $SU_3$ irrep, namely 60% $[(4, 10)_\pi (18, 4)_\nu](22, 14)$ for $^{162}$Dy. The second excited $K = 0$ state is a strong mixture of two $SU_3$-irreps, 47% of $[(10, 4)_\pi (20, 0)_\nu](30, 4)$ and 34% of $[(10, 4)_\pi (18, 4)_\nu](28, 8)$. As before, only a couple irreps contribute to more than about 2% to the eigenvectors.

The $SU_3$ content of the eigenvectors is fairly constant across the states within a band. The percentages vary slowly and smoothly from that of the band-head as one moves up the band to states with higher values of the angular momentum. To see this, the $SU_3$ content of calculated eigenvectors for the states in the ground-state band in $^{162}$Dy are given in Table 3. Only the basis states that contribute to more than 2% are identified.

| $J_\#$ | Th. | $(\lambda_\pi, \mu_\pi)$ | $(\lambda_\nu, \mu_\nu)$ | $(\lambda, \mu)$ |
|---|---|---|---|---|
| $0_1$ | 59.3 | ( 10, 4) | ( 18, 4) | ( 28, 8) |
|  | 6.5 | ( 10, 4) | ( 18, 4) | ( 30, 4) |
|  | 20.1 | ( 10, 4) | ( 20, 0) | ( 30, 4) |
|  | 7.1 | ( 12, 0) | ( 18, 4) | ( 30, 4) |
|  | 2.7 | ( 10, 4) | ( 18, 4) | ( 32, 0) |
|  | 3.0 | ( 12, 0) | ( 20, 0) | ( 32, 0) |
| $0_a$ | 91.8 | ( 4, 10) | ( 18, 4) | ( 22, 14) |
|  | 4.2 | ( 10, 4) | ( 20, 0) | ( 30, 4) |
|  | - | ( 10, 4) | ( 18, 4) | ( 28, 8) |
|  | - | ( 10, 4) | ( 20, 0) | ( 30, 4) |
|  | - | ( 12, 0) | ( 18, 4) | ( 30, 4) |
| $0_b$ | 33.3 | ( 10, 4) | ( 18, 4) | ( 28, 8) |
|  | 47.0 | ( 10, 4) | ( 20, 0) | ( 30, 4) |
|  | 11.0 | ( 12, 0) | ( 20, 0) | ( 32, 0) |
|  | 5.9 | ( 4, 10) | ( 18, 4) | ( 22, 14) |
|  | - | ( 12, 0) | ( 18, 4) | ( 30, 4) |
| $2_\gamma$ | 81.4 | ( 10, 4) | ( 18, 4) | ( 28, 8) |
|  | 5.2 | ( 10, 4) | ( 20, 0) | ( 30, 4) |
|  | 4.6 | ( 12, 0) | ( 18, 4) | ( 30, 4) |
|  | 4.5 | ( 4, 10) | ( 18, 4) | ( 22, 14) |

Table 2. $SU_3$ content of calculated eigenvectors for states of the four low-lying band-heads in $^{162}$Dy. The percentage distributions of each eigenvector across the $(\lambda, \mu)$ values are given in the second column. Only basis states that contribute to more than 2% are identified.

| $(\lambda_\pi, \mu_\pi)$ | $(\lambda_\nu, \mu_\nu)$ | $(\lambda, \mu)$ | 0 | 2 | 4 | 6 | 8 |
|---|---|---|---|---|---|---|---|
| ( 10, 4) | ( 18, 4) | ( 28, 8) | 59.3 | 59.3 | 59.4 | 59.6 | 61.9 |
| ( 10, 4) | ( 20, 0) | ( 30, 4) | 20.1 | 19.5 | 18.2 | 16.3 | 13.9 |
| ( 10, 4) | ( 18, 4) | (30, 4) | 6.5 | 6.2 | 5.7 | 4.9 | 4.0 |
| ( 12, 0) | ( 18, 4) | ( 30, 4) | 7.1 | 6.9 | 6.7 | 6.3 | 5.7 |
| ( 10, 4) | ( 18, 4) | ( 32, 0) | 2.7 | 2.6 | 2.3 | - | - |
| ( 12, 0) | ( 20, 0) | ( 32, 0) | 3.0 | 2.9 | 2.8 | 2.5 | 2.1 |
| ( 10, 4) | ( 16, 5) | ( 26, 9) | - | - | - | 3.2 | 3.5 |

Table 3. $SU_3$ content of calculated eigenvectors for the states in the ground-state band in $^{162}$Dy using parameters from Table 1. Only the basis states that contribute to more than 2% are identified.

### 3.1. B(E2) transitions

Theoretical and experimental [10] $B(E2)$ transitions strengths between the states in the ground state band in $^{162}$Dy are shown in Table 4. The agreement between the calculated and experimental numbers is excellent. The $B(E2; 2_1 \to 4_1)$ is equal to within 1% of the experimental value, and the last two calculated $B(E2)$ values differ from the experimental values by less than 0.1 $e^2b^2$ which is well within the experimental error. Excellent agreement with experimental $B(E2)$ data is also observed in $^{162}$Dy and $^{164}$Dy. Contributions to the quadrupole moments from the nucleons in the unique parity orbitals are parameterized through an effective charge [9], $e_f$, with $e_\nu = e_f$, and $e_\pi = 1 + e_f$, so the $E2$ operator is given by [9]

$$Q_\mu = e_\pi Q_\pi + e_\nu Q_\nu . \qquad (8)$$

Theoretical intra-band B(E2) transition strengths between the states in the $K = 2$ as well as the first and second excited $K = 0$ bands are given in Table 5. Note that the strengths of the transitions probabilities are consistent across all four bands (Tables 4 and 5).

| $J_i \to J_f$ | $B(E2; J_i \to J_f)$ ($e^2b^2$) | |
|---|---|---|
|  | Exp. | Theory |
| $0_1 \to 2_1$ | $5.134 \pm 0.155$ | 5.134 |
| $2_1 \to 4_1$ | $2.675 \pm 0.102$ | 2.635 |
| $4_1 \to 6_1$ | $2.236 \pm 0.127$ | 2.325 |
| $6_1 \to 8_1$ | $2.341 \pm 0.115$ | 2.201 |

Table 4. Experimental and theoretical $B(E2)$ transition strengths between members of ground state band of $^{162}$Dy.

### 3.2. M1 transitions and the $K = 1_1^+$ band

Another test for the theory is the $M1$ transition strength distributions that can be obtained using eigenvectors of the diagonalized Hamiltonian (1). The calculated and experimental $M1$ strength distributions for the Dy nuclei are given in Figs. 1 b), d) and f). The calculated $M1$ strength distribution for $^{168}$Er is given in Fig. 2 b). For illustrative purposes, the energies and M1 transition spectra are given opposite one another.

The starting point for a geometric interpretation of the scissors mode within the framework of the $SU_3$ shell model is the well-known relation of the $SU_3$ symmetry group to $Rot(3)$, the symmetry group of the rotor [11, 12]. The structure of the intrinsic Hamiltonian allows for a rotor-model interpretation of the coupled $SU_3$ irreps $(\lambda_\pi, \mu_\pi)$ and $(\lambda_\nu, \mu_\nu)$ for protons and neutrons, respectively. According to the Littlewood rules [13] for coupling Young diagrams, the allowed product configuration can be expressed in mathematical terms by using three integers (m, l, k):

$$(\lambda_\pi, \mu_\pi) \otimes (\lambda_\nu, \mu_\nu) = \oplus_{m,l,k}(\lambda_\pi + \lambda_\nu - 2m + l, \mu_\pi + \mu_\nu - 2l + m)^k, \quad (9)$$

where the parameters $l$ and $m$ are defined in a fixed range given by the values of the initial $SU_3$ representations. In this formulation, $k$ serves to distinguish between multiple occurrences of equivalent $(\lambda, \mu)$ irreps in the tensor product. The number of $k$ values is equal to the outer multiplicity, $\rho_{max}$ ($k = 1, 2, \ldots, \rho_{max}$). The $l$ and $m$ labels can be identified [15] with excitation quanta of a two-dimensional oscillator involving relative rotations ($\theta$, the angle between the principal axes of the proton and neutron system, and $\phi$, the angle between semi-axes of the proton and neutron system) of the proton-neutron system,

$$m = n_\theta, l = n_\phi. \tag{10}$$

| $K = 2$ | $2_\gamma \to 3_\gamma$ | 2.480 |
|---|---|---|
|  | $2_\gamma \to 4_\gamma$ | 1.060 |
|  | $3_\gamma \to 4_\gamma$ | 1.630 |
|  | $4_\gamma \to 5_\gamma$ | 1.145 |
|  | $4_\gamma \to 6_\gamma$ | 1.625 |
|  | $5_\gamma \to 6_\gamma$ | 0.716 |
|  | $6_\gamma \to 7_\gamma$ | 0.607 |
|  | $6_\gamma \to 8_\gamma$ | 1.685 |
| $K = 0_2$ | $0_a \to 2_a$ | 4.193 |
|  | $2_a \to 4_a$ | 2.272 |
|  | $4_a \to 6_a$ | 2.153 |
|  | $6_a \to 8_a$ | 2.175 |
| $K = 0_3$ | $0_b \to 2_b$ | 3.517 |
|  | $2_6 \to 4_b$ | 1.901 |
|  | $4_b \to 6_b$ | 2.017 |
|  | $6_b \to 8_b$ | 2.030 |

Table 5. Theoretical $B(E2)$ transition strengths between states of the $K = 2$, $K = 0_2$, and $K = 0_3$ bands of $^{162}$Dy. The energies are labeled with the subindex $\gamma$ for the $K = 2$ band, $a$, and $b$ for the first and second excited $K = 0$ bands.

| Nucleus | $\sum B(M1)[\mu_N^2]$ | | |
|---|---|---|---|
|  | Experiment | Calculated | |
|  |  | Pure $SU_3$ | Theory |
| $^{160}$Dy | 2.48 | 4.24 | 2.32 |
| $^{162}$Dy | 3.29 | 4.24 | 2.29 |
| $^{164}$Dy | 5.63 | 4.36 | 3.05 |

Table 6. Total B(M1) strength from experiment [10] and the present calculation.

These correspond to two distinct type of $1^+$ motion, the scissors and twist modes, and their realization in terms of the pseudo-$SU_3$ model.

The $SU_3$ irreps obtained from the tensor product (9) that contain a $J^\pi = 1^+$ state are those corresponding to $(m, l, k) = (1, 0, 1)$, $(0, 1, 1)$, $(1, 1, 1)$, and $(1, 1, 2)$. A pure $SU_3$ picture gives rise of a maximum of four $1^+$ states that are associated with the scissors, twist, and double degenerate scissors-plus-twist modes [(1,1,1) and (1,1,2)] [15]. Results for the Dy isotopes, assuming a pure pseudo-$SU_3$ scheme, are given in Table 7.

The experimental results [10] given in Figs. 1 b), d), and f) suggest a much larger number of $1^+$ states with non-zero $M1$ transition probabilities from the $0^+$ ground state. The $SU_3$ breaking residual interactions lead to a fragmentation in the $M1$ strength distribution, since the ground state $0^+$ is in that case a combination of several $SU_3$ irreps, each with allow M1 transitions to other $SU_3$ irreps. Overall, the total $M1$ strength is in reasonable agreement with the experimental results (Table 6). In $^{164}$Dy the total $M1$ strength is slightly underestimated, which may be due to spin admixtures in the wavefunction, which is not included in this work.

| Nucleus | $[(\lambda_\pi, \mu_\pi)$ | $(\lambda_\pi, \mu_\pi)$ | $(\lambda, \mu)]_{gs}$ | $(\lambda, \mu)_{1^+}$ | B(M1) | mode |
|---|---|---|---|---|---|---|
| $^{160-162}$Dy | (10,4) | (18,4) | (28,8) | ( 29, 6) | 0.56 | t |
| | | | | ( 26, 9) | 1.77 | s |
| | | | | $(27, 7)^1$ | 1.82 | s+t |
| | | | | $(27, 7)^2$ | 0.083 | t+s |
| $^{164}$Dy | (10,4) | (20,4) | (30,8) | (31,6) | 0.56 | t |
| | | | | (28,9) | 1.83 | s |
| | | | | (29,7) | 1.88 | s+t |
| | | | | (29,7) | 0.09 | t+s |

Table 7. B(M1) transition strengths $[\mu_N^2]$ in the pure symmetry limit of the pseudo $SU_3$ model. The strong coupled pseudo-$SU_3$ irrep $(\lambda, \mu)_{gs}$ for the ground state is given with its proton and neutron sub-irreps and the irreps associated with the $1^+$ states, $(\lambda', \mu')_{1^+}$. In addition, each transition is labeled as a scissors (s) or twist (t) or combination mode.

According to Eq. (5), and depending upon the values of the $(\lambda, \mu)$ irreps, there are several $K = 1$ bands. Since we are interested in $M1$ transitions from the ground state to the $1^+$ states, several $(\lambda, \mu)$ that allow $K = 1$ states are included in the configuration space. As a result, the model predicts several $K = 1$ bands. The first calculated $K^\pi = 1^+$ state lies very close to the first experimental one (the numbers are identical in the first three digits) in $^{160}$Dy and $^{162}$Dy. Moreover this being the band-head of a $K^\pi = 1^+$ band, the band is also remarkable well described. The calculated and experimental energies of the $K^\pi = 1^+$ band in $^{162}$Dy are given in Table 8 together with the B(E2) transition strengths. A few additional states of angular momentum 5, 6, 7, and 8 are predicted as members of the first $K = 1$ band in $^{162}$Dy.

| $J_i \to J_f$ | $B(E2; J_i \to J_f)$ $(e^2b^2)$ | Energy($J_f$) [MeV] Exp. | Th. |
|---|---|---|---|
|  |  | 1.746 | 1.749 |
| $1_1 \to 2_5$ | 1.312 | 1.783 | 1.797 |
| $2_5 \to 3_3$ | 0.944 | 1.840 | 1.905 |
| $2_5 \to 4_5$ | 1.170 | 1.904 | 1.941 |
| $4_5 \to 5_3$ | 0.802 | - | 1.993 |
| $4_5 \to 6_4$ | 0.648 | - | 2.066 |
| $6_4 \to 7_2$ | 1.291 | - | 2.150 |
| $6_4 \to 8_4$ | 1.203 | - | 2.292 |

Table 8. The experimental and calculated energies of the $K = 1$ band in $^{162}$Dy using parameters from table 1. The B(E2) transition probabilities are given in the second column. The energy of the $J = 1$ state is given in the first row.

### 3.3. $K^\pi = 4_1^+$ band

In the pure $SU_3$ symmetry limit there are no interaction terms that mix states of different $(\lambda, \mu)$ irreps. The energies of the nonzero band-heads associated with nonzero $K$ values are then fixed, by-in-large by the $b$ parameter (though band mixing within an irrep is allowed this is normally small, vanishing in the limit of large prolate configurations). A simple exercise yields a value of $2^2 \times b$ for the $K = 2$, $J = 2$ state and a value of $4^2 \times b$ for the $K = 4$, $J = 4$ state. If, for example we assume the $b$ value from Table 1 for $^{160}$Dy, the values for these two states would be 0.4 MeV for $J_{K=2_2}$ and 1.6 MeV for $J_{K=4_4}$, respectively. However in the present version of the model these states are mixed due primarily to the single-particle energies and pairing interactions. In the present calculations $b$ parameter adjusted to give a best fit to the band-head of the $K = 2$ state only, yielding a value of 2.29 MeV for the band-head of the $K = 4$ band in $^{162}$Dy, which is only slightly higher than the experimental value.

In these nuclei there is an experimentally known low-lying $K^\pi = 4^+$

| $J_i \to J_f$ | Energy($J_f$) [MeV] | | $B(E2; J_i \to J_f)$ |
|---|---|---|---|
| | Exp. | Th. | ($e^2 b^2$) |
| $4_6$ | 1.536 | 1.977 | |
| $4_6 \to 5_4$ | 1.634 | 2.168 | 0.360 |
| $4_6 \to 6_6$ | 1.752 | 2.209 | 0.976 |
| $6_6 \to 7_4$ | 1.888 | 2.514 | 0.125 |
| $6_6 \to 8_6$ | - | 2.527 | 1.784 |

Table 9. Experimental and calculated energies of states of the angular momentum indicated by the right-side entry of the first column for members of the $K = 4$ band of $^{162}$Dy. The B(E2) transition probabilities are given in the last column. The energy of the $J = 4$ band-head is given in the first row.

| $J_i \to J_f$ | Energy($J_f$) [MeV] | | $B(E2; J_i \to J_f)$ |
|---|---|---|---|
| $4_5$ | 2.173 | 1.836 | |
| $4_5 \to 5_4$ | - | 1.944 | 1.147 |
| $4_5 \to 6_7$ | - | 2.124 | 0.493 |

Table 10. Experimental and calculated energies of states of the indicated angular momentum given by the right-side entry of the first column for members of the $K = 4$ band in $^{164}$Dy. The B(E2) transition probabilities are given in the last column. The energy of the $J = 4$ band-head is given in the first row.

band. A comparison between the experimental and calculated energies are given in Table 9 together with B(E2) transition probabilities for $^{162}$Dy. One additional state of total angular momentum 8 is identified as a member of the $K = 4$ band, by both a strong $B(E2)$ transition to the state of angular momentum 6 and through similar $SU_3$ content with the other states in this band. In the $^{164}$Dy case there is an experimentally known $K = 4$ state at $2.173 MeV$. The theory predicts a $K = 4$ state at lower energy ($1.836 MeV$) and moreover according to the theory it should be a band-head.

## 4. Conclusions

This study of $^{160,162,164}$Dy and $^{168}$Er shows that pseudo-spin zero neutron and proton configuration with a relatively few pseudo-$SU_3$ irreps with largest $C_2$ values suffices to obtain good agreement with known experimental results. The Hamiltonian that was used included single-particle energies, the quadrupole-quadrupole interaction, and neutron and proton pairing terms, all with strengths fixed by systematics, plus four smaller rotor-like terms with strengths that were varied to maximize agreement with observations.

A consistent set of 'free' parameters was obtained. The results generated extended beyond quantities that were used in the fitting procedure, including intra-band B(E2) strengths, the M1 strength distribution of the ground state, and band-head energies of the first $K^\pi = 1^+$ and $K^\pi = 4^+$ bands.

The M1 strength distributions were not fit to the data. Nevertheless, in all cases the summed strength was found to be in good agreement with the experiment numbers. The pseudo-$SU_3$ model therefore offers a microscopic shell-model interpretation of the "scissors" mode [14], and in addition, it reveals a "twist" degree of freedom that corresponds to allowed relative angular motion of the proton and/or neutron sub-distribution [15]. By adding one-body and two-body pairing interactions to the Hamiltonian, it was possible to describe the experimentally observed fragmentation of the M1 strength. The results suggest that more detailed microscopic description of other properties of heavy deformed nuclei, such as g-factors and beta decay, may finally be within reach of a bona fide microscopic theory.

Supported in part by the U.S. National Science Foundation through a regular grant, PHY-9970769, and a Cooperative Agreement, EPS-9720652, that includes matching from the Louisiana Board of Regents Support Fund.